# Coupled valence and spin state transition in $(Pr_{0.7}Sm_{0.3})_{0.7}Ca_{0.3}CoO_3$

F. Guillou[1], Q. Zhang[1], Z. Hu[2], C. Y. Kuo[2], Y.Y. Chin[3], H.J. Lin[3], C. T. Chen[3], A. Tanaka[4], L.H. Tjeng[2], and V. Hardy[1]

[1] *Laboratoire CRISMAT, ENSICAEN, UMR 6508 CNRS, 6 Boulevard du Maréchal Juin, 14050 Caen Cedex, France.*

[2] *Max Planck Institute for Chemical Physics of Solids, Nöthnizer Straße 40, 01187 Dresden, Germany*

[3] *National Synchrotron Radiation Research Center (NSRRC), 101 Hsin-Ann Road, Hsinchu 30077, Taiwan*

[4] *Department of Quantum Matters, ADSM, Hiroshima University, Higashi-Hiroshima, 739-8530, Japan*

**Abstract**

The coupled valence and spin state transition (VSST) taking place in $(Pr_{0.7}Sm_{0.3})_{0.7}Ca_{0.3}CoO_3$ was investigated by soft x-ray absorption spectroscopy (XAS) experiments carried out at the Pr-$M_{4,5}$, Co-$L_{2,3}$, and O-$1s$ edges. This VSST is found to be composed of a sharp Pr/Co valence and Co spin state transition centered at $T^* \sim 89.3$ K, followed by a smoother Co spin-state evolution at higher temperatures. At $T < T^*$, we found that the praseodymium displays a mixed valence $Pr^{3+}/Pr^{4+}$ with about 0.13 $Pr^{4+}$/f.u., while all the $Co^{3+}$ is in the low-spin (LS) state. At $T \sim T^*$, the sharp valence transition converts all the $Pr^{4+}$ to $Pr^{3+}$ with a corresponding $Co^{3+}$ to $Co^{4+}$ compensation. This is accompanied by an equally sharp spin state transition of the $Co^{3+}$ from the low to an incoherent mixture of low and high spin (HS) states. An involvement of the intermediate spin (IS) state can be discarded for the $Co^{3+}$. While above $T^*$ and at high temperatures the system shares rather similar properties as Sr-doped $LaCoO_3$, at low temperatures it behaves much more like $EuCoO3$ with its highly stable LS configuration for the $Co^{3+}$. Apparently, the mechanism responsible for the formation of $Pr^{4+}$ at low temperatures also helps to stabilize the $Co^{3+}$ in the LS configuration despite the presence of $Co^{4+}$ ions. We also found out that that the $Co^{4+}$ is in an IS state over the entire temperature range investigated in this study (10-290 K). The presence of $Co^{3+}$ HS and $Co^{4+}$ IS at elevated temperatures facilitates the conductivity of the material.

# I. INTRODUCTION

For some transition-metal ions placed in certain ligand field symmetries, the two main energy terms acting on the intra-atomic electronic configuration (i.e., the crystal-field splitting, $\Delta_{CF}$, and the Hund exchange, $J_H$) turn out to be very close to each other. This can result in the stabilization of different electronic configurations, as well as the occurrence of transitions between them, induced by temperature or external stimuli like pressure or light-irradiation. In condensed matter, such a peculiar situation is most often encountered for $3d^5$ and $3d^6$ cations in a cubic-like environment.

The Spin State Transitions (SST) are among the most intriguing phenomena in solid state physics, and they have been the subject of continuing interest since more than fifty years. In oxides, the cation most prone to such a behavior is $Co^{3+}$ ($3d^6$) located at an octahedral site, which can exhibit not only low-spin (LS) and high-spin (HS) states (corresponding to $t_{2g}^6 e_g^0$ and $t_{2g}^4 e_g^2$, respectively), but also the so-called intermediate-spin state (IS), when the ligand hole state is important (formally noted $t_{2g}^5 e_g^1$, even though it is expected to be a highly hybridized state). [1-3]

To investigate the SST associated to $Co^{3+}$, the archetypical material which has been intensively studied since the 1950s is the perovskite $LaCoO_3$.[4-7] It exhibits two broad transitions, one located around $T_1 \sim 100$ K and the second around $T_2 \sim 500$ K. In spite of a huge amount of experimental and theoretical works, there is no consensus about the nature of the $Co^{3+}$ spin states associated to these transitions.[8-13] Regarding $T_1$ for instance, there is still an intense debate in which the two most often investigated pictures involve a transition from a LS state at low-temperature to either an IS state [8,14,15] or a mixed LS/HS state.[10,11,13,16] Even though the former scenario is the prevailing one nowadays, some authors strongly dispute it and even question the possibility of an IS state for $Co^{3+}$ in $LaCoO_3$.[11] It must be emphasized that the thermally-driven transitions in $LaCoO_3$ are quite smooth, rather looking like crossovers associated with gradual populations of excited states, a feature which add a degree of complexity in all interpretations.

In 2002, Tsubouchi et al. reported in $Pr_{0.5}Ca_{0.5}CoO_3$ another type of SST,[17] which is much sharper, located at a characteristic temperature that is hereafter referred to as $T^*$. It was found that the transition taking place at $T^* \sim 90$ K in $Pr_{0.5}Ca_{0.5}CoO_3$ is actually a first-order transition, associated to discontinuous changes in various physical quantities like magnetization, entropy and resistivity. As compared to $LaCoO_3$, another basic difference is that $Pr_{0.5}Ca_{0.5}CoO_3$ is a mixed valence compound (0.5 $Co^{3+}$ and 0.5 $Co^{4+}$ per formula unit when considering $Pr^{3+}$ and $Ca^{2+}$). It thus deserves to be noted that $Co^{4+}$ ($3d^5$) sitting at an octahedral site is also susceptible to take different spin states. One could even add that, in the original paper reporting on the possibility of IS states in cobalt oxides, the occurrence of such a state was claimed to be more favorable for $Co^{4+}$ than for $Co^{3+}$.[1] Soon after the discovery of a SST in $Pr_{0.5}Ca_{0.5}CoO_3$, there was a systematic investigation of perovskite compounds ($ABO_3$), with Co at the octahedral B site and various combinations of rare-earth and alkaline-earth elements at the A site. It was found that a sharp SST can take place over a quite wide range of composition, which however appeared to always require the presence of Pr and Ca on the A site of the perovskite, leading to the formulation $(Pr_{1-y}Ln_y)_{1-x}Ca_xCoO_3$.[18-24]

The interpretation of the SST at $T^*$ initially given by Tsubouchi et al. was to consider a transition from $Co^{3+}$ LS to $Co^{3+}$ IS upon warming, while $Co^{4+}$ stay in a LS state at all temperatures.[17,19] Starting from 2010 however, an accumulation of data pointed to a more complex scenario, involving a valence transition in praseodymium. First, in a neutron powder diffraction study of $Pr_{0.5}Ca_{0.5}CoO_3$, Barón-González et al. [25] observed that the Co-O bonds are almost unaltered at $T^*$, whereas the Pr-O ones change a lot, a result leading them to suggest the possibility of a charge transfer between Co and Pr at the transition. This viewpoint was supported soon after by electronic structure calculations performed by Knížek et al.,[26] who reported a partial change from ($Pr^{4+}/Co^{3+}$) below $T^*$ to ($Pr^{3+}/Co^{4+}$) above $T^*$. Then, analyzing heat capacity data in $(Pr_{1-y}Y_y)_{0.7}Ca_{0.3}CoO_3$ compounds, Hejtmánek et al.[27] found a Schottky-like anomaly at $T < T^*$, a feature strongly suggesting the presence of the Kramers ion $Pr^{4+}$ ($4f^1$). Finally, in 2011, experiments using soft x-ray absorption spectroscopy (XAS) provided direct evidences of the presence of $Pr^{4+}$ below $T^*$. In $Pr_{0.5}Ca_{0.5}CoO_3$, García-Muñoz et al. [28] and Herrero-Martín et al. [29] both demonstrated a partial oxidation of $Pr^{3+}$ into $Pr^{4+}$ when crossing $T^*$ upon cooling. The quantitative estimate of the fraction of $Pr^{4+}$ at $T << T^*$, however, was found to vary between 0.075 and 0.13 $Pr^{4+}$/f.u..[28,29] Moreover, it was emphasized that this phenomenon is more complex than a simple valence change, since it also involves the development of a strong hybridization between Pr *4f* and O *2p* orbitals below $T^*$.[29]

The most generally accepted scheme for the valence and spin state transition (VSST) discussed above is the following:[17,19,21,22,24] Below $T^*$, a fraction of Pr is in a nominal $Pr^{4+}$ state, the rest being in the more usual $Pr^{3+}$ state; both the $Co^{3+}$ and $Co^{4+}$ are in a LS state; When crossing $T^*$ upon warming, $Pr^{4+}$ transform to $Pr^{3+}$, this being counterbalanced by a concomitant increase of the $Co^{4+}$ content at the expense of the $Co^{3+}$ one; In terms of spin state, the $Co^{3+}$ LS transform to IS while the $Co^{4+}$ remain in a LS state. While the existence of a valence change involving Pr at the transition is well documented now, the nature of the Co spin states below and above the transition remains a matter of debate. As a matter of act, a very recent work favored the transition towards LS/HS for $Co^{3+}$ above $T^*$, but it did not categorically rule out the IS scenario [30]. Moreover, we emphasize that the LS state of the $Co^{4+}$ was assumed rather than demonstrated in most of the previous studies, which can be considered as a highly optimistic approach since $Co^{4+}$ is a cation which is susceptible to exhibit various spin states [1].

In this context, we reinvestigated the VSST in a $(Pr_{1-y}Ln_y)_{1-x}Ca_xCoO_3$ compound with the objective to identify in detail the nature of the spin states of not only the $Co^{3+}$ but also of the $Co^{4+}$ ions. We have carried out soft x-ray absorption (XAS) experiments at the Pr-$M_{4,5}$, Co-$L_{2,3}$, and O-*1s* edges and made a detailed analysis using a combination of well-defined reference spectra and multiplet calculations. We made an effort to perform the experiments at many closely spaced temperature intervals, from well below to above $T^*$, in order to determine how much of the spin state transition is coupled to the Pr valence transition and how much it is driven by temperature as in $LaCoO_3$. This study was carried out on $(Pr_{0.7}Sm_{0.3})_{0.7}Ca_{0.3}CoO_3$, a choice motivated by two main advantages as compared to $Pr_{0.5}Ca_{0.5}CoO_3$: first, an easier synthesis procedure;[22] second, a larger content of $Co^{3+}$ which is favorable to investigate the controversial issue of its spin state above $T^*$. Note also that $(Pr_{0.7}Sm_{0.3})_{0.7}Ca_{0.3}CoO_3$ exhibits a VSST located at $T^* \sim 90$ K (like $Pr_{0.5}Ca_{0.5}CoO_3$ in most of the previous studies) and a lot of its physical properties are well documented in the literature.[21,22]

## II. EXPERIMENTAL DETAILS

Polycrystalline samples of $(Pr_{0.7}Sm_{0.3})_{0.7}Ca_{0.3}CoO_3$ (whose formula can also be written $Pr_{0.49}Sm_{0.21}Ca_{0.3}CoO_3$) were prepared by solid-state reaction using stoichiometric proportions of $Pr_6O_{11}$, $Sm_2O_3$, $CaO$ and $Co_3O_4$. The powders were intimately ground and the mixture was pelletized in form of 2×2×10 mm bars. These bars were then sintered at 1200 °C in flowing oxygen for 36 h. To ensure good oxygen stoichiometry, the samples were finally annealed in high-pressure (130 bar) $O_2$ atmosphere for 48 h at 600 °C. Powder x-ray diffraction patterns attested to the purity of the material which shows a single-phase with orthorhombic symmetry (space group P*nma*) and parameters yielding a unit cell volume of 215.99(5) Å$^3$ in accord with the literature.[21]

Electrical resistivity ($\rho$), magnetization (*M*) and heat capacity (*C*) measurements were carried out in a commercial Physical Properties Measurements System (PPMS): resistivity was measured by a standard four-probe method, magnetization by an extraction method, and heat capacity by a relaxation method supplemented by a "Single Pulse" technique.[31] These data were collected upon warming, in a magnetic field equal to zero for $\rho$ and C, and equal to 1 T for *M*. The magnetic data can be presented in the form of *dc* susceptibility $\chi = M/H$, since the *M(H)* curves were found to exhibit a well linear behaviour over the temperature range of interest in the present study (*T* > 20 K).

The temperature dependent XAS spectra at the Co-$L_{2,3}$, O-*1s* and Pr-$M_{4,5}$ edges were measured at the BL11A beamline of National Synchrotron Radiation Research Center (NSRRC) in Taiwan. The Co-$L_{2,3}$ and Pr-$M_{4,5}$ spectra were taken in the total electron yield (TEY) mode, while the O-*K* spectra were collected in the fluorescence yield (FY) mode, with a photon energy resolution of 0.3 and 0.15 eV, respectively. Spectra were recorded at eight temperatures between 10 K and 290 K, with a 10 K spacing around the transition. The pellet samples were cleaved *in situ* in an ultrahigh-vacuum chamber with pressure in the $10^{-10}$ mbar range. NiO and CoO single crystals were measured *simultaneously* to serve as energy references for the O-*K* and Co-$L_{2,3}$ edges, respectively.

# III. RESULTS AND DISCUSSION

Figure 1 shows a prominent transition close to 90 K on the $\chi(T)$, $\rho(T)$ and $C(T)$ curves. The value of this transition temperature, as well as the concomitant decrease in resistivity and increase in susceptibility that is observed upon warming, are features in line with previous reports on this compound.[21,22] In the present study, the sharpness of the transition is also supported by the heat capacity data, which exhibits a remarkably high peak centred at $T^* \sim 89.3$ K, with a full-width at half-maximum that does not exceed 1 K [Fig. 1(d)].

## A. Valence transition in Pr

Let us first address the change in the valence state of Pr when crossing $T^*$, by considering the evolution with temperature of the Pr-$M_{4,5}$ XAS spectrum. In Fig. 2, one clearly observes the appearance of a broad shoulder at the high-energy side of the $M_4$ and $M_5$ peaks when cooling below T*, a feature which can be attributed to an increase in the average valence of Pr, i.e. in our case, a partial transformation of $Pr^{3+}$ towards $Pr^{4+}$. The enlargement of the $M_5$ high-energy foot displayed in the inset shows a very abrupt change as a function of temperature, the transition taking place between 80 and 90 K, consistently with the $T^*$ value.[32] In contrast, the temperature dependence of the Pr spectra, both below and above $T^*$, is found to be very weak.

Let us now quantify the amount of tetravalent praseodymium below $T^*$ by considering that the valence state of Pr in this regime is a mixture between $Pr^{3+}$ and $Pr^{4+}$. Assuming that the spectrum at 290 K is representative of $Pr^{3+}$ only, as done in the previous studies, the evaluation of the $Pr^{4+}$ content at 10 K was derived as follows: first, the 290 K spectrum multiplied by a factor < 1 was subtracted from the 10 K spectrum till reaching a "difference spectrum" showing the typical features of pure $Pr^{4+}$, as observed in the measurements of $PrO_2$ (e.g., ratio close to 4 between the main peak and the high energy shoulder at $M_5$);[33,34] Then, the fraction of $Pr^{4+}$ present at 10 K is associated to the spectral weight of this difference spectrum in the total one (at the same temperature of 10 K). Doing so, the estimated $Pr^{4+}$ content was found to be 0.13 $Pr^{4+}$/f.u..

Our compound containing also samarium which is another rare-earth susceptible to exhibit different valence states (e.g., $Sm^{2+}$), we investigated the Sm-$M_{4,5}$ edges between 1060 and 1130 eV as a function of temperature. We observed in all cases a spectrum well compatible with $Sm^{3+}$ as reported in $Sm_2O_3$,[35] without any change in temperature, demonstrating that Sm does not take part in the VSST process.

## B. Valence and spin state transitions in Co

To probe the valence and spin states of the Co ions, we considered the Co-$L_{2,3}$ and O-$K$ absorption edges. It can be noted that the Co ($2p \rightarrow 3d$) transitions at the Co-$L_{2,3}$ edge involve directly the relevant valence shell and are extremely sensitive to the charge and spin states [11,36-40].

Figure 3 shows the evolution of the XAS spectrum as a function of temperature, between 10 and 290 K. One observes that these spectra closely resemble those collected in $LaCoO_3$,[6,11] the most significant difference being a relative larger intensity of the shoulders located just above the main peaks at $L_3$ and $L_2$ (noted C and E), which can be attributed to the presence of $Co^{4+}$ in $(Pr_{0.7}Sm_{0.3})_{0.7}Ca_{0.3}CoO_3$. There is also a striking similarity in the temperature dependence of this spectrum, which manifests itself by two main features: (i) at low temperatures, the XAS spectra taken below 80 K in $(Pr_{0.7}Sm_{0.3})_{0.7}Ca_{0.3}CoO_3$ clearly look like that of $LaCoO_3$ at 20 K [11], showing a narrow peak at ~796 eV at the $L_2$ edge (noted D), which is a feature known to be the hallmark of $Co^{3+}$ LS; (ii) at high temperatures, the XAS spectra above 90 K in $(Pr_{0.7}Sm_{0.3})_{0.7}Ca_{0.3}CoO_3$ exhibit a low energy shoulder at ~780 eV (noted A) below the main peak of the Co-$L_3$ edge (noted B), which is identical to what is observed in $LaCoO_3$ above ~ 600 K [6,11]. This feature is typical of the presence of $Co^{3+}$ in a spin state higher than LS, and is expected to be more pronounced for HS than for IS. [11,36]

This similarity between $LaCoO_3$ and $(Pr_{0.7}Sm_{0.3})_{0.7}Ca_{0.3}CoO_3$ in the temperature dependence of the Co-$L_{2,3}$ spectrum indicates that a change in $Co^{3+}$ from a LS state to a higher spin state does take place in the latter compound upon arising temperature. There is, however, a notable difference in the *shape* of this temperature dependence, since $(Pr_{0.7}Sm_{0.3})_{0.7}Ca_{0.3}CoO_3$ exhibits a rapid change around $T^*$, contrary to $LaCoO_3$ in which one observes only a smooth transition spread out over a wide temperature range; Looking in more detail at the data of $(Pr_{0.7}Sm_{0.3})_{0.7}Ca_{0.3}CoO_3$ (bottom panels of Fig. 3), one observes that the spectral change is very small from 10 K to 80 K, before exhibiting an abrupt variation between 80 K to 90 K, and finally a weaker –but well visible- temperature dependence up to 290 K.

Figure 4 displays the temperature dependent O-$K$ spectrum, which is also sensitive to a spin state transition in $Co^{3+}$ [6,15,36]. Here, we are mainly interested in the lowest lying states, *i.e.*, the pre-edge region lying below 532 eV, which corresponds to transitions from the O *1s* core level to the O *2p* orbitals mixed with unoccupied Co *3d* states [6,36,39,40]. As the temperature is increased, one observes a sudden transfer of spectral weight from the higher energy peak at ~ 529.1 eV to the lower energy peak at ~ 528.2 eV. Such a temperature-induced transfer –also present in $LaCoO_3$ (between 420 and 550 K)[6] and $Ba_2Co_9O_{14}$ [41]– is an additional support to the presence of a spin state transition involving $Co^{3+}$ in $(Pr_{0.7}Sm_{0.3})_{0.7}Ca_{0.3}CoO_3$. Indeed, the electron promotion from $t_{2g}$ and $e_g$ orbitals that characterizes such a SST is expected to yield this type of spectral weight transfer. A striking feature of Fig. 4 is the steepness of the spectral weight transfer, that is found to be much sharper than in $LaCoO_3$ [6]. It can also be noted that the spectral change observed between 80 K to 90 K in $(Pr_{0.7}Sm_{0.3})_{0.7}Ca_{0.3}CoO_3$ is even more pronounced at the O *1s* edge (Fig. 4) than at the Co-$L_{2,3}$ edge (Fig. 3). This can be ascribed to the fact that the spectral weight below 532 eV at the O-$K$ edge is expected to be affected not only by the spin state transition of $Co^{3+}$ [6] but also by the valence change in Pr; indeed $Pr^{4+}$ has a clear signature in this energy range due to a strong 4*f*/O*2p* covalence, whereas $Pr^{3+}$ has not [42].

After having established the existence of a SST in the Co ions, the problem is now to determine the nature of the spin states of $Co^{3+}$ and $Co^{4+}$ that are involved in this transition. We note that the only issue on which there is a general consensus so far is that $Co^{3+}$ is in a pure LS state at $T \ll T^*$. Before addressing the delicate issue of the $Co^{3+}$ spin state above $T^*$, let us start with the case of $Co^{4+}$ which is another important, but often overlooked issue.

*B1. Spin state of $Co^{4+}$*

In most of the previous studies on VSST in cobaltites, it was assumed that the $Co^{4+}$ are in a LS state, and that they stay in this state over the whole *T*-range investigated so far, i.e. 300 K. However, we previously noted that the possibility of higher spin state should be considered for $Co^{4+}$, in particular the intermediate spin state. Even though the archetypical example of $Co^{4+}$ IS remains the perovskite SrCoO$_3$ [1], the occurrence of this state was recently reported in other oxides such as Sr$_2$CoO$_4$ [43] and Ca$_3$Co$_4$O$_9$.[44] On the other hand, it is widely admitted that Co$^{4+}$ is LS in BaCoO$_3$ [45]. From a structural viewpoint, we emphasize that (Pr$_{0.7}$Sm$_{0.3}$)$_{0.7}$Ca$_{0.3}$CoO$_3$ is much closer to SrCoO$_3$ (cubic perovskite with corner-shared CoO$_6$ octahedra) than to BaCoO$_3$ (hexagonal 2H-structure showing chains of face-sharing CoO$_6$), which is a first indication pointing to the possibility of a IS state in our material.

To investigate this issue more precisely, one can compare Co-$L_{2,3}$ spectra at 10 K to those calculated considering experimental reference spectra for Co$^{3+}$ LS (EuCoO$_3$),[37] Co$^{4+}$ LS (BaCoO$_3$) [38] and Co$^{4+}$ IS (SrCoO$_3$).[1] In this analysis, we considered the ratio Co$^{4+}$/Co$^{3+}$ as a free parameter. Figure 5 shows the best fitting obtained with each of these two spin states of Co$^{4+}$, and it is clear that one obtains better agreement with Co$^{4+}$ IS. Moreover, the content of Co$^{4+}$ derived from this fitting is found to be 0.16 per f.u., i.e., well consistent with the value 0.17 (= 0.30 - 0.13) that is expected from the previously estimated amount of Pr$^{4+}$ (0.13/f.u.). Therefore, it appears that IS is the most likely spin state of Co$^{4+}$ below $T^*$. In addition, since the Co$^{4+}$ HS state is supposed to lie at much higher energy [20] -not reachable till room temperature- we will consider that the Co$^{4+}$ ion remains in this IS state over the whole temperature range of the present study.

*B2. Spin state of Co$^{3+}$*

For Co$^{3+}$, one has to face the problem of the absence of undisputed reference compounds for the IS and HS spin-states in an octahedral environment. This prevents us from directly comparing the data to composite spectra made of experimental reference spectra, as done above for Co$^{4+}$. Therefore, to address the nature of the Co$^{3+}$ spin state, one has to shift to another strategy where the experimental temperature dependent Co-$L_{2,3}$ XAS spectra are compared to theoretical simulations. In the present study, such calculated curves were derived from a configuration-interaction (CI) cluster model, including full-atomic multiplet theory, crystal-field effect and hybridization with the O $2p$ ligands.[46-48] The validity of this approach is supported by its proven ability to reproduce the experimental XAS spectra of reference compounds, i.e. containing cations for which there is no doubt about the valence and spin states. Limiting ourselves to the case of $Co^{3+}/Co^{4+}$, such compounds are $BaCoO_3$ (for $Co^{4+}$ LS),[38] $SrCoO_3$ (for $Co^{4+}$ IS),[1] $EuCoO_3$ (for $Co^{3+}$ LS),[36] as well as $Sr_2CoO_3Cl$ (for $Co^{3+}$ HS in a pyramidal environment).[36] Using theoretical spectra calculated by the CI cluster model was also found to be a fruitful approach to address the nature of the Co spin states in various compounds where this issue was under question, like for instance in $La_{1.5}Sr_{0.5}CoO_4$ and $Na_xCoO_2$.[38-39] The main parameters of the CI cluster model are the O *2p* to Co *3d* charge-transfer energy (Δ), the *3d-3d* and *3d-2p* Coulomb energies ($U_{dd}$ and $U_{cd}$, respectively), the ionic part of the crystal-field (10*Dq*), and the O *2p* to Co *3d* transfer integral involving the $e_g$ orbitals (*pdσ*). It must be emphasized that the actual crystal-field splitting is significantly larger than 10*Dq* owing to the ligand field contribution (covalency effects) associated to the hybridization with the O *2p* orbitals.

The values of parameters used to calculate the theoretical spectra of $Co^{4+}$ IS, $Co^{3+}$ HS and $Co^{3+}$ LS in the present study are given in Ref. 49. They are similar to those previously reported in the literature,[1,11,36,37] and are also well consistent with general trends expected when varying the valence or spin-state of a *3d* cation. As a matter of fact, the main expected change between the LS and HS states of $Co^{3+}$ is a smaller crystal-field splitting, which is presently accounted for by a decrease in both 10*Dq* and |*pdσ*|, the latter effect being actually predominant.[6,11] About the influence of the valence state, the main effect when going from +3 to +4 for late transition metals is a marked decrease in Δ, which can yield negative values in the case of Co.[47-48] Our values are also consistent with the relationship $\Delta^{n+} \approx \Delta^{(n-1)+} - U_{dd}$, previously reported for the couples $Fe^{3+}/Fe^{2+}$ and $Co^{4+}/Co^{3+}$.[38,50]

To quantify hereafter the percentages of $Co^{3+}$ (LS and HS) and $Co^{4+}$ (IS) in the case of $(Pr_{0.7}Sm_{0.3})_{0.7}Ca_{0.3}CoO_3$, the contribution of 2% of $Co^{2+}$ (using experimental spectra from CoO) was first subtracted from the experimental Co-$L_{2,3}$ XAS spectra shown in Fig. 6. Note that the presence of such a level of $Co^{2+}$ impurity is often observed in cobaltites, and is generally ascribed to a slight under stoichiometry of oxygen.

The calculation procedure was first tested with the 10 K spectrum, for which the population of all the present Co species can be anticipated, as discussed in Section B.1. As displayed in Fig. 6(a), it turns out that the calculated spectrum obtained by combining the sum of 83 % $Co^{3+}$ LS and 17% $Co^{4+}$ IS with a edge jump curve is in perfect agreement with the experimental data. We emphasize that the matching to the data is as good as that obtained when combining experimental spectra of reference compounds (top panel of Fig. 5), a feature which confirms the reliability of the simulated spectra derived from the CI cluster model

When the temperature is increased above $T^*$, let us first consider that the spin state transition of $Co^{3+}$ leads to a mixture between LS and HS spin states. As previously done in $LaCoO_3$, the temperature dependence of the Co-$L_{2,3}$ spectrum was analyzed by using an incoherent sum of the LS and HS contributions for $Co^{3+}$.[11] This was combined with a contribution from $Co^{4+}$ IS, whose content was found to increase abruptly (by ~ 10%) when crossing $T^*$. Doing so, we were able to obtain good simulations of the spectra at all temperatures, as exemplified in Fig. 6(b) and (c) for 90 K and 290 K, respectively.

Although the results of Fig. 6 are well consistent with the achievement of a mixed LS/HS state for $Co^{3+}$ above $T^*$, let us consider the alternative possibility of a $Co^{3+}$ IS state. To do so, we considered the spectrum at room temperature. We first calculated the $Co^{3+}$ IS spectrum (blue line) that is displayed in Fig.6(d). Note that, in the cubic symmetry, the $Co^{3+}$ IS states lie too high in energy and cannot be stabilized; therefore, the $Co^{3+}$ IS spectrum is not taken from the ground states, but rather from high-lying excited states. In order to stabilize the $Co^{3+}$ IS ground state, one has to introduce a huge $e_g$ splitting (more than 2 eV), via a Jahn-Teller distortion, in the full multiplet calculation. Then, we compared the experimental spectrum taken at 290 K with the sum of theoretical spectra for $Co^{3+}$ IS and $Co^{4+}$ IS, keeping the ratio between them as an adjustable parameter. It turns out that even the best simulation shown in Fig. 6(d) exhibits a poor overlapping with the experimental data, in marked contrast with the almost perfect superimposition shown by the LS/HS scenario in Fig. 6(c). Accordingly, one can consider that these results provide us with a direct evidence against the occurrence of a (LS → IS) transition in the VSST of $(Pr_{1-y}Ln_y)_{1-x}Ca_xCoO_3$ compounds.

In the literature, the onset of metalliclike conductivity above $T^*$ is often used as an argument to support the (LS→IS) scenario, since easy electronic delocalization is expected between $Co^{3+}$ IS and $Co^{4+}$ LS. It must be emphasized, however, that our picture characterized by $Co^{3+}$ HS and $Co^{4+}$ IS yields a very similar situation from the viewpoint of electronic mobility.[51] In both cases indeed, one is dealing with two cations having the same core spin at the $t_{2g}$ triplet and that just differ by one electron on the $e_g$ doublet, a situation allowing to interchange the states by moving only one $e_g$ electron. Accordingly, we note that the couple (Co$^{3+}$ HS/Co$^{4+}$ IS) considered in the present work is as compatible as (Co$^{3+}$ IS/Co$^{4+}$ LS) with the good electronic conductivity that is observed above $T^*$.

### C. Temperature dependence of the transition

Simulations of the Co-$L_{2,3}$ XAS spectra were carried out for all of the experimental spectra at the temperatures measured, using combinations of contributions from the Co$^{4+}$ IS, Co$^{3+}$ LS and Co$^{3+}$ HS. In Fig. 7(a), we reported the percentages of each of these species, that lead to the best agreement with the data. In Fig. 7(b) is also plotted the temperature variation of the Pr$^{3+}$ content derived from the Pr-$M_{4,5}$ spectra using the procedure described in Section III.A. One observes a steep increase in Pr$^{3+}$ at $T^*$, with a much weaker temperature dependence both below and above the transition. This overall variation turns out to be very close to that previously derived in Pr$_{0.5}$Ca$_{0.5}$CoO$_3$ [28,29] and (Pr$_{1-y}$Y$_y$)$_{0.7}$Ca$_{0.3}$CoO$_3$ (y=0.075 and 0.15) [52] using different methods of analysis of XAS data.

In Fig. 7(a), we can observe that the Co$^{3+}$ LS content is reduced from 83 % to 55% between 10 K to 90 K, and is further reduced to 32 % at 290 K, while the Co$^{3+}$ HS content is increased from 0% to 16% and then to 37%. The Co$^{4+}$ fraction increases from 17% to 29% between 10 K and 90 K, and then remains essentially constant up to 290 K. Consistently with the charge balance, the variation of Pr$^{3+}$ in Fig. 7(b) exhibits a profile similar to that of Co$^{4+}$. Quantitatively speaking, the Pr$^{3+}$ content per f.u. increases from 0.36 to 0.44 between 10 K and 90 K, and reaches 0.47 at 100 K, i.e., a value closely approaching that of a pure trivalent state (0.49 per f.u.).

Figure 7 shows that the temperature dependences of the Co$^{3+}$ LS, Co$^{3+}$ HS, and Co$^{4+}$ IS fractions between 10 K and 90 K are all very similar to that of the Pr$^{4+}$/Pr$^{3+}$ valence change, at least within the accuracy of the experiment and analysis. Following a low-temperature regime showing a smooth evolution (which must be regarded cautiously since these variations are actually within the experimental uncertainty), the main feature found in all cases is a large change centered at $T^*$, i.e., between 80 K and 90 K.

The overall behavior displayed in Fig. 7 thus indicates that the spin state transition of Co$^{3+}$ in the 10 K to 90 K temperature range is intimately coupled to the Pr$^{4+}$/Pr$^{3+}$ valence change. A different behavior sets in at higher temperatures, i.e., at $T > T^*$, where one can observe that the temperature dependence of the $Co^{3+}$ HS and LS contents is significantly more pronounced than that of $Pr^{3+}$ and $Co^{4+}$, the latter species being actually already close to their saturation values. Considering the variations between 100 K and 290 K, the $Pr^{3+}$ and $Co^{4+}$ contents only change by less than 5%, whereas the population of $Co^{3+}$ HS is increased by more than 66%.

In the high temperature regime ($T > T^*$), it deserves to be noted that $(Pr_{0.7}Sm_{0.3})_{0.7}Ca_{0.3}CoO_3$ exhibits properties similar to those of Sr-doped $LaCoO_3$ [53]: a substantial fraction of the $Co^{3+}$ is in the HS state, the spin state transition is gradual, and the resistivity is no longer that of a semiconductor but rather that of a bad metal. The low temperature behavior of $(Pr_{0.7}Sm_{0.3})_{0.7}Ca_{0.3}CoO_3$ is however, very different from that of Sr-doped $LaCoO_3$ or even from that of undoped $LaCoO_3$. Despite the high percentage of $Co^{4+}$ (~17%), all of the $Co^{3+}$ remains in the LS state. Increasing the temperature to 40K or even 60 K, the amount of $Co^{3+}$ HS remains minute (< 2%), much smaller than in $LaCoO_3$. In this respect, $(Pr_{0.7}Sm_{0.3})_{0.7}Ca_{0.3}CoO_3$ is more similar to $EuCoO_3$ which has a very stable $Co^{3+}$ LS configuration over a wide range of temperatures. Apparently, the mechanism responsible for the formation of $Pr^{4+}$ at low temperatures somehow also prevents the $Co^{4+}$ ions to interact with the $Co^{3+}$ thereby leaving the $Co^{3+}$ in a stable LS configuration.

## IV. CONCLUSION

We investigated the valence and spin state transition (VSST) in $(Pr_{0.7}Sm_{0.3})_{0.7}Ca_{0.3}CoO_3$, whose most salient manifestation is a sharp first-order transition at $T^* \sim 89.3$ K, visible in all physical properties. Using XAS measurements at the Co-$L_{2,3}$, Pr-$M_{4,5}$ and O-$K$ edges, we found evidence for valence state transitions in Co and Pr in the vicinity of $T^*$. The spin state transition in $Co^{3+}$ follows this valence transition from low temperature up to $T^*$. Above $T^*$, we essentially observe a gradual continuation of the spin state transition of the $Co^{3+}$. At low temperatures, we observed that $Co^{3+}$ and $Co^{4+}$ are in LS and IS states, respectively, while Pr has a mixed valence state $Pr^{3+}/Pr^{4+}$ with about 0.13 $Pr^{4+}$/fu. Crossing $T^*$ upon warming, the transition exhibits a jump, associated with the stabilization of ions having larger ionic radius, i.e., $Co^{3+}$ HS and $Pr^{3+}$ at the expense of $Co^{3+}$ LS and $Pr^{4+}$, a behavior that is consistent with the sudden volume expansion previously observed in related compounds.[17,25,28] Our analysis of the XAS spectra indicates that the spin-state transition of the $Co^{3+}$ involves the LS and HS states, and that the IS state does not play a role. Contrary to the literature, we found that the spin state of $Co^{4+}$ is IS instead of LS. The presence of the IS for the $Co^{4+}$ together with the HS for the $Co^{3+}$ leaves open the possibility for a good electronic conduction as is observed at elevated temperatures. Obviously, the issue of the formation of $Pr^{4+}$ and the role of the bonding with the oxygens [54] will deserve further investigation both from theoretical and experimental viewpoints as this seems to be a key ingredient of the VSST.

## ACKNOWLEDGMENTS

This work was supported by the European program "SOPRANO" under Marie Curie actions (Grant. No. PITN-GA- 2008-214040).

**Figures captions**

Fig. 1: (a) Temperature dependence of the resistivity in zero-field ; (b) Temperature dependence of the *dc* susceptibility( measured in 1 T); (c) Temperature dependence of the heat capacity in zero-field ; (d) Enlargement of (c) around the peak.

Fig. 2 Temperature dependent Pr-$M_{4,5}$ XAS spectrum. The arrows displayed close to the main features of the spectrum indicate the direction of increasing temperature. The inset is an enlargement of the high-energy foot of $M_5$.

Fig. 3: Temperature dependent Co-$L_{2,3}$ XAS spectrum. Labels A, B, C, D, E mark features discussed in the text. In each case, the direction of the arrow corresponds to decreasing temperature. Lower panels show enlargements of the energy ranges corresponding to the A and D features.

Fig. 4 : Temperature dependence of the pre-edge region of the O-*K* XAS spectrum.

Fig. 5: Experimental Co-$L_{2,3}$ XAS spectrum at 10 K compared to simulations using *experimental reference spectra* for $Co^{3+}$ ($EuCoO_3$) [from Ref. 37] combined with either $Co^{4+}$ IS ($SrCoO_3$) [from Ref. 1] or $Co^{4+}$ LS ($BaCoO_3$) [from Ref. 38].

Fig. 6: Experimental (black circles) Co-$L_{2,3}$ XAS spectra compared to simulations (red lines) at various temperatures: (a) 10 K; (b) 90 K; (c,d) 290 K. The simulations combine theoretical spectra for $Co^{3+}$ LS (black), $Co^{3+}$ HS (magenta), $Co^{4+}$ IS (green), as well as $Co^{3+}$ IS (blue) in (d). The dashed lines represent edge jump in all cases. The label A in (a) marks the same feature as in Fig. 3.

Fig. 7: (a) Temperature dependence of the populations of $Co^{3+}$ LS, $Co^{3+}$ HS and $Co^{4+}$ IS derived from the simulations of the experimental Co-$L_{2,3}$ XAS spectra; (b) temperature dependence of the $Pr^{3+}$ content extracted from the Pr-$M_{4,5}$ XAS spectra.

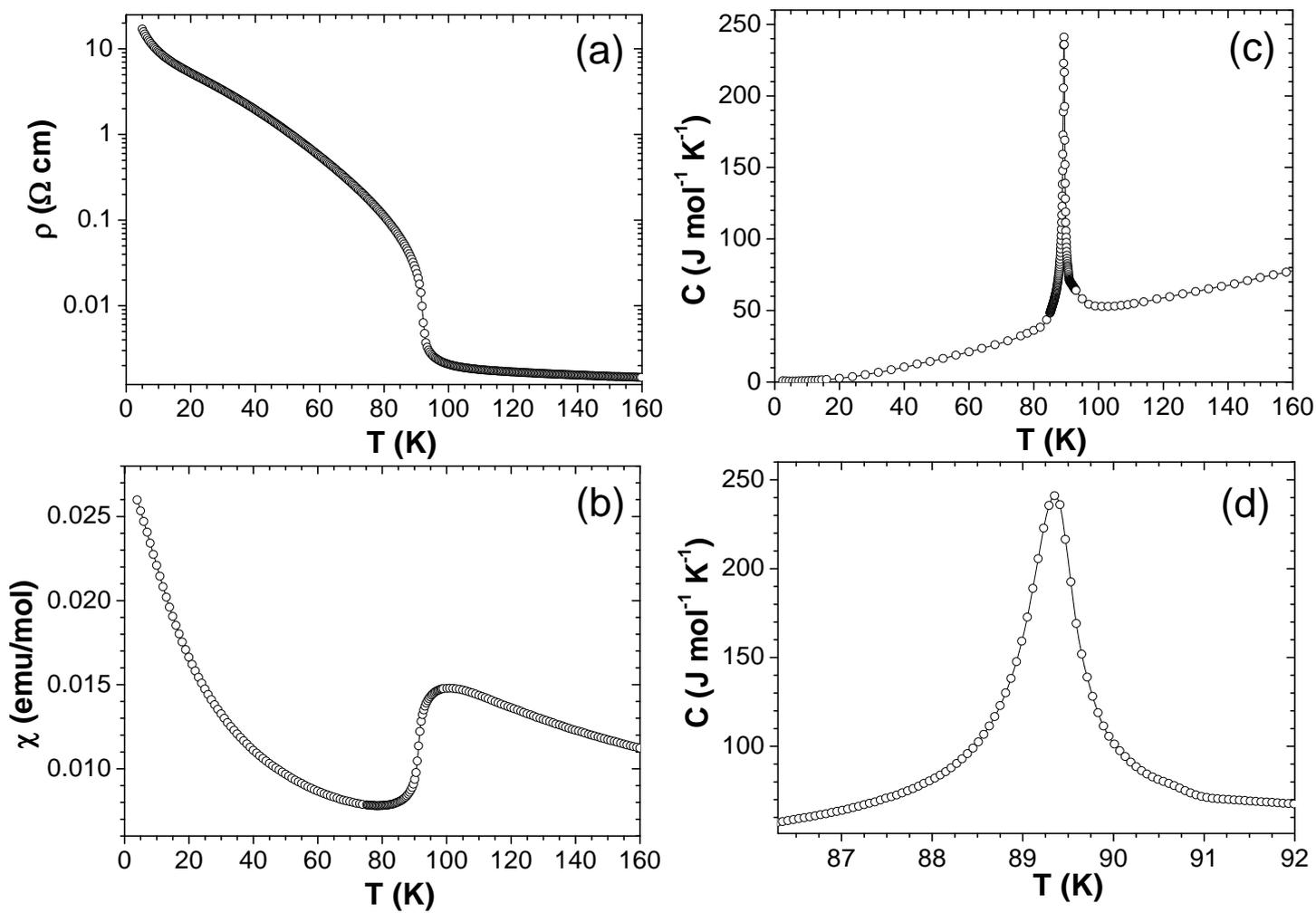

Figure 1

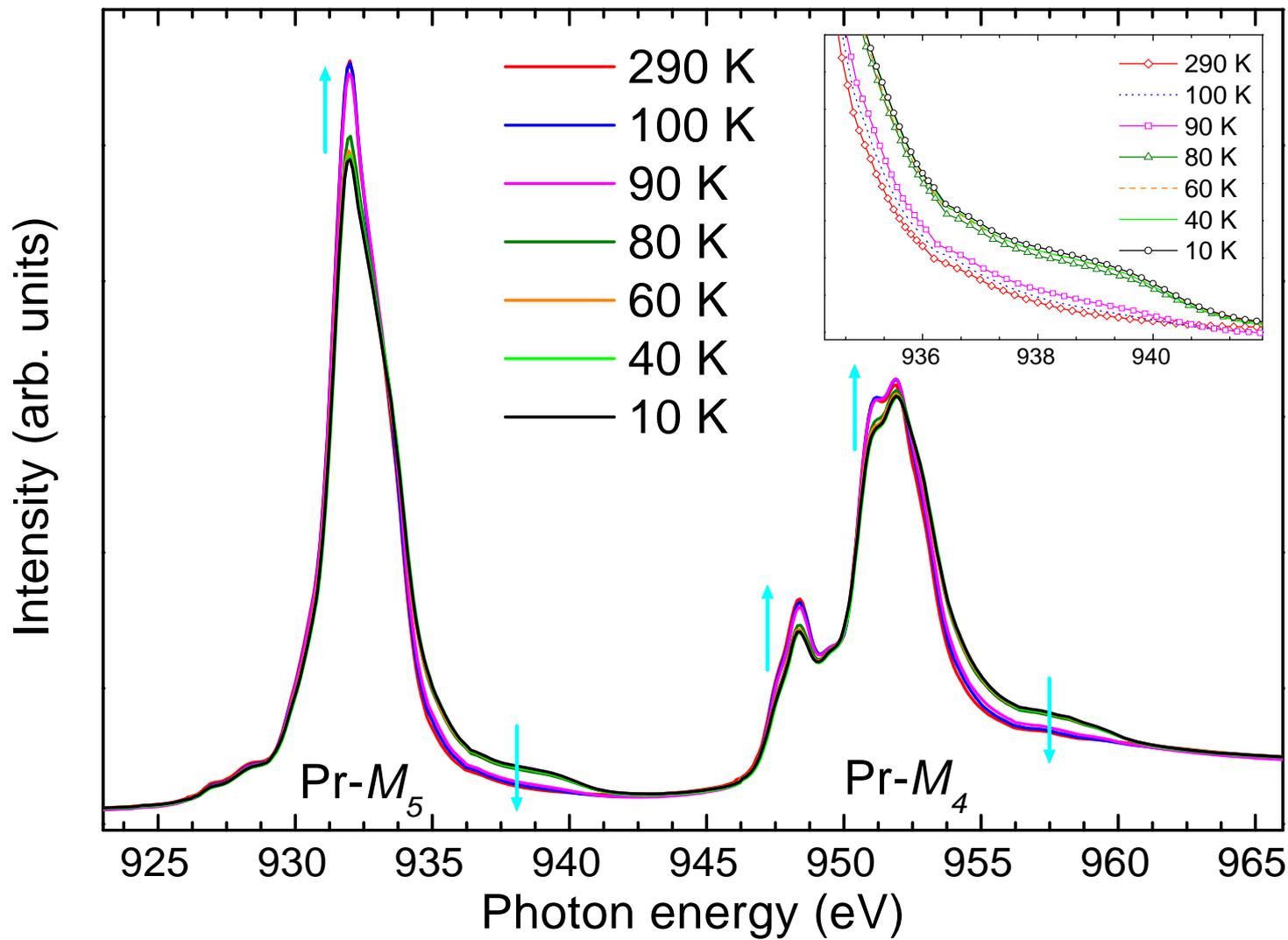

Figure 2

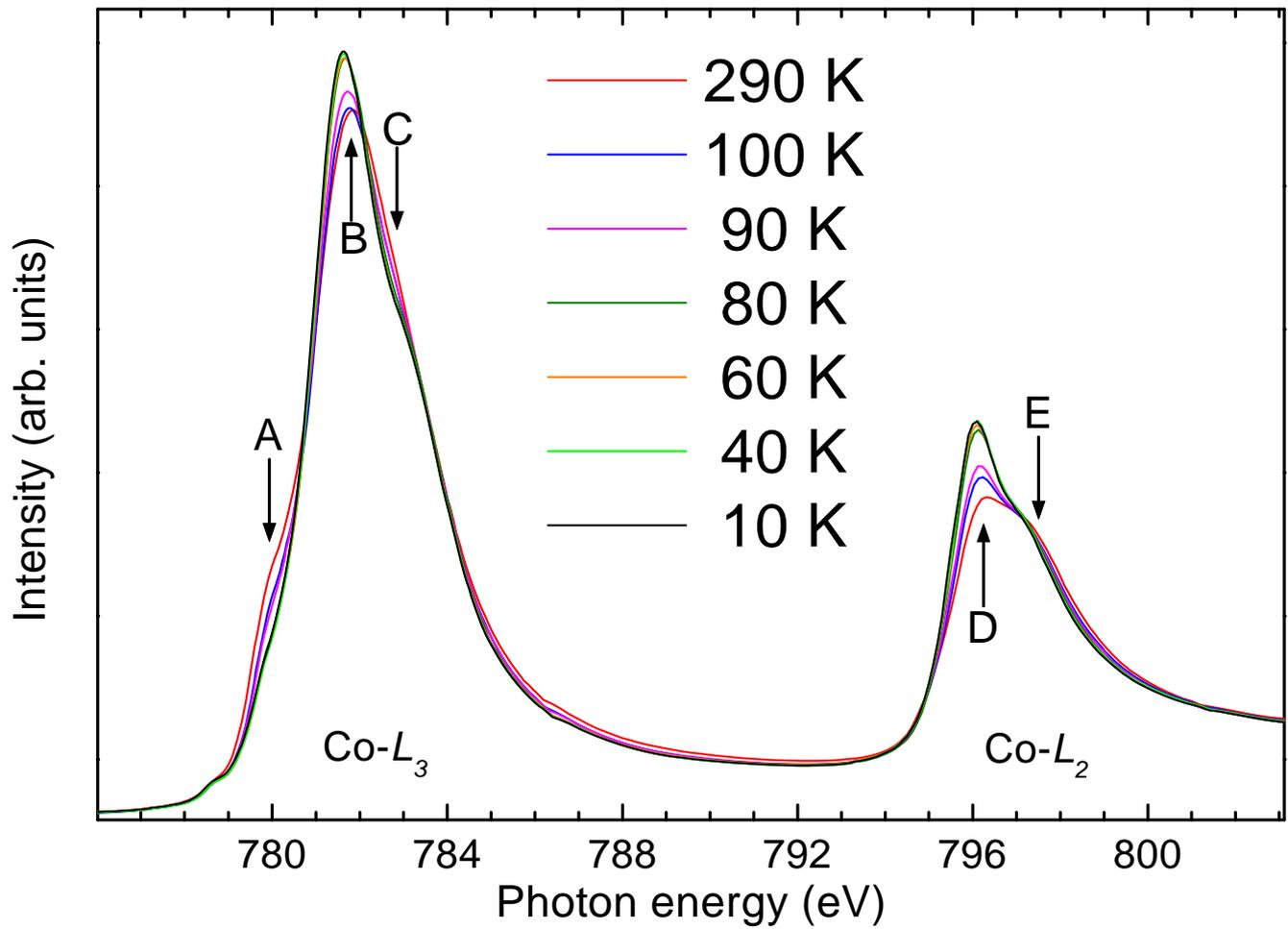

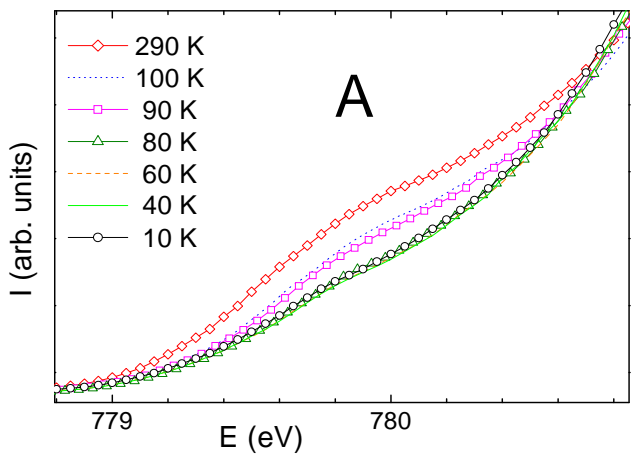

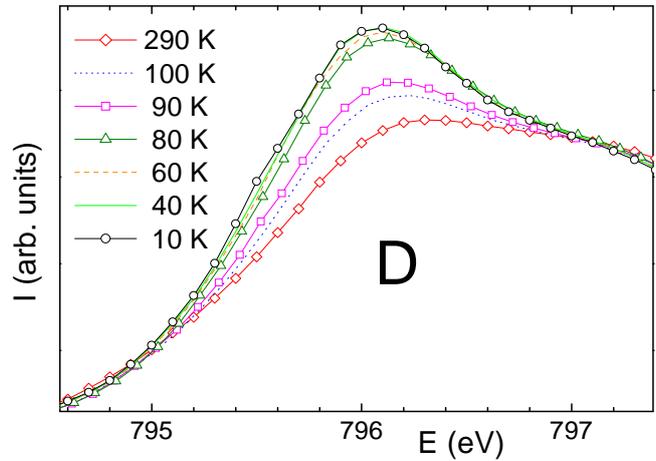

Fig 3

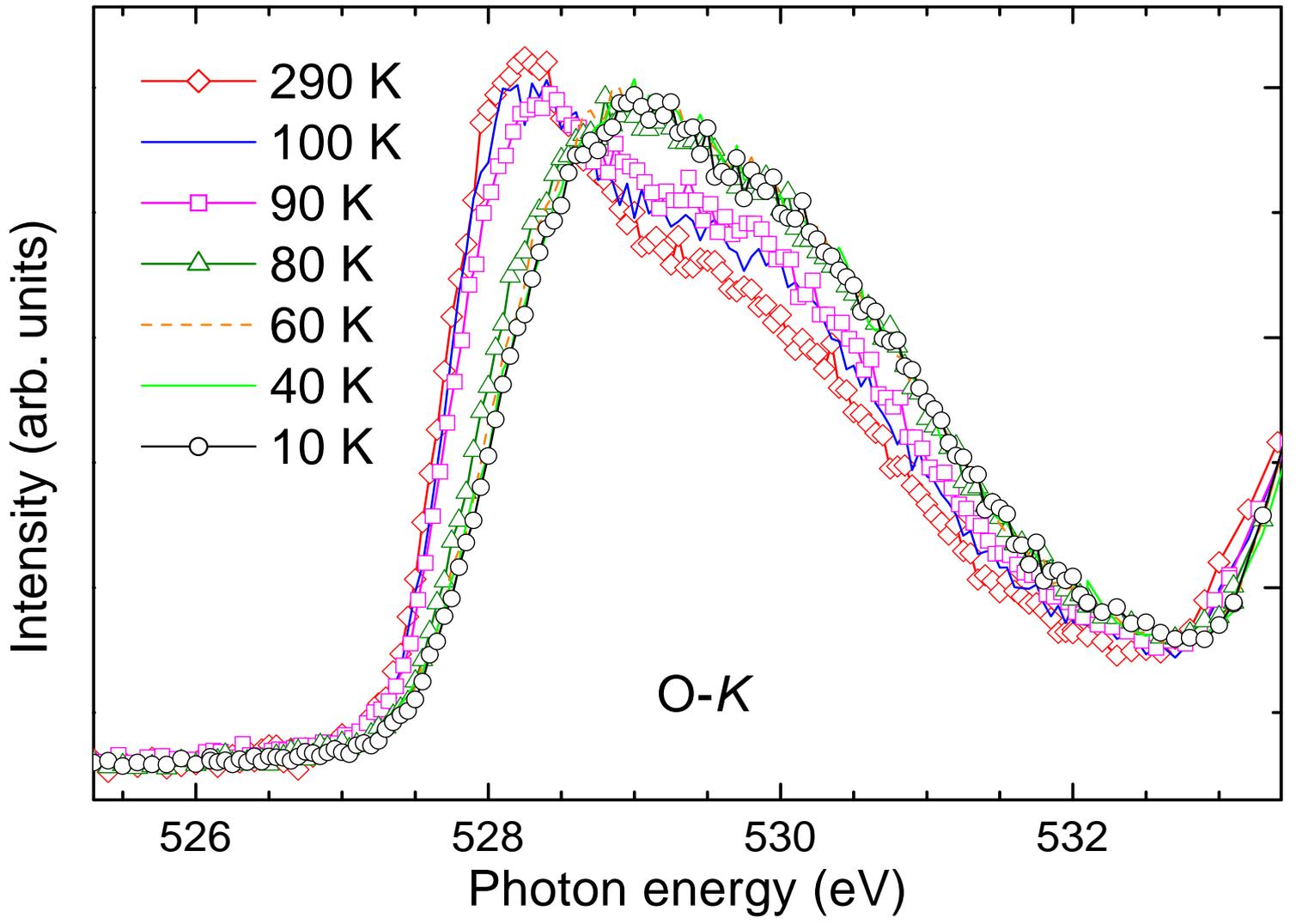

Figure 4

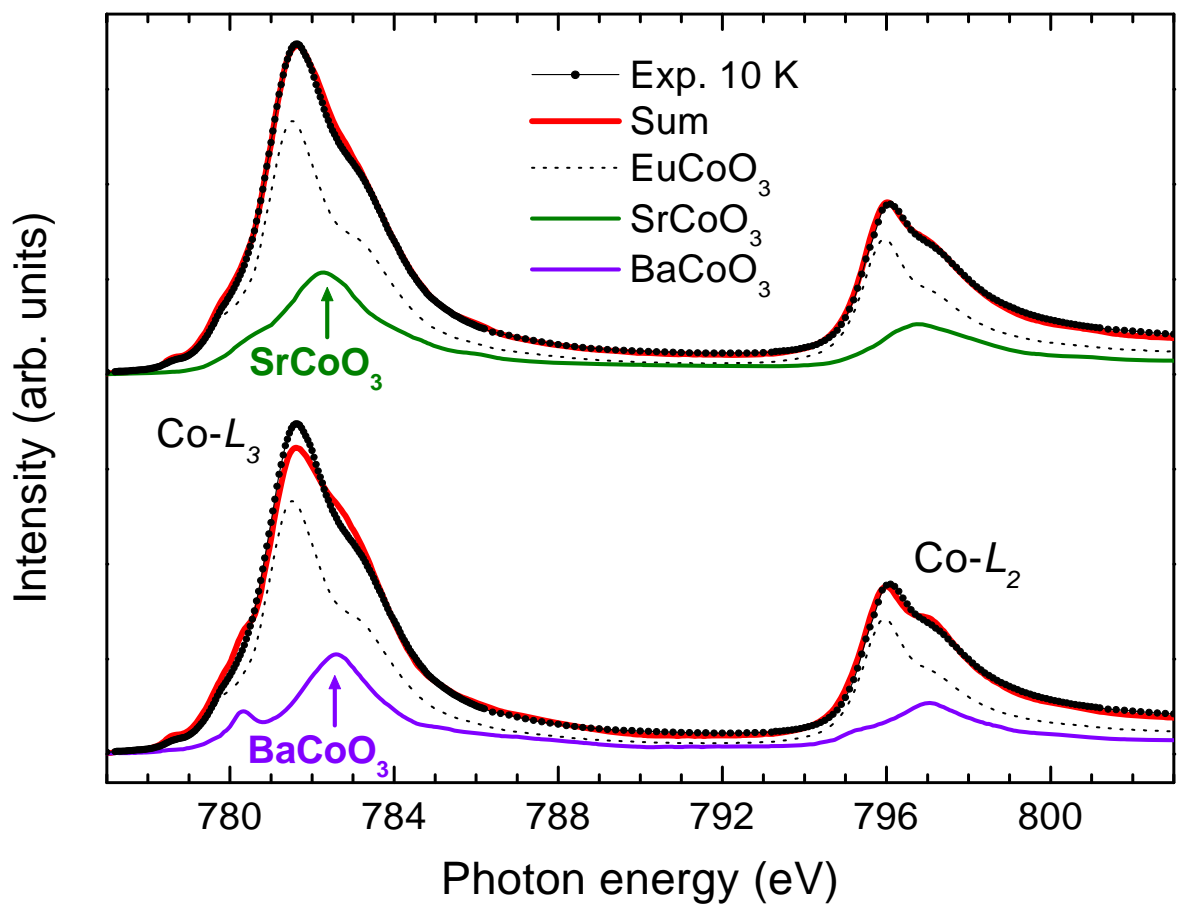

Figure 5

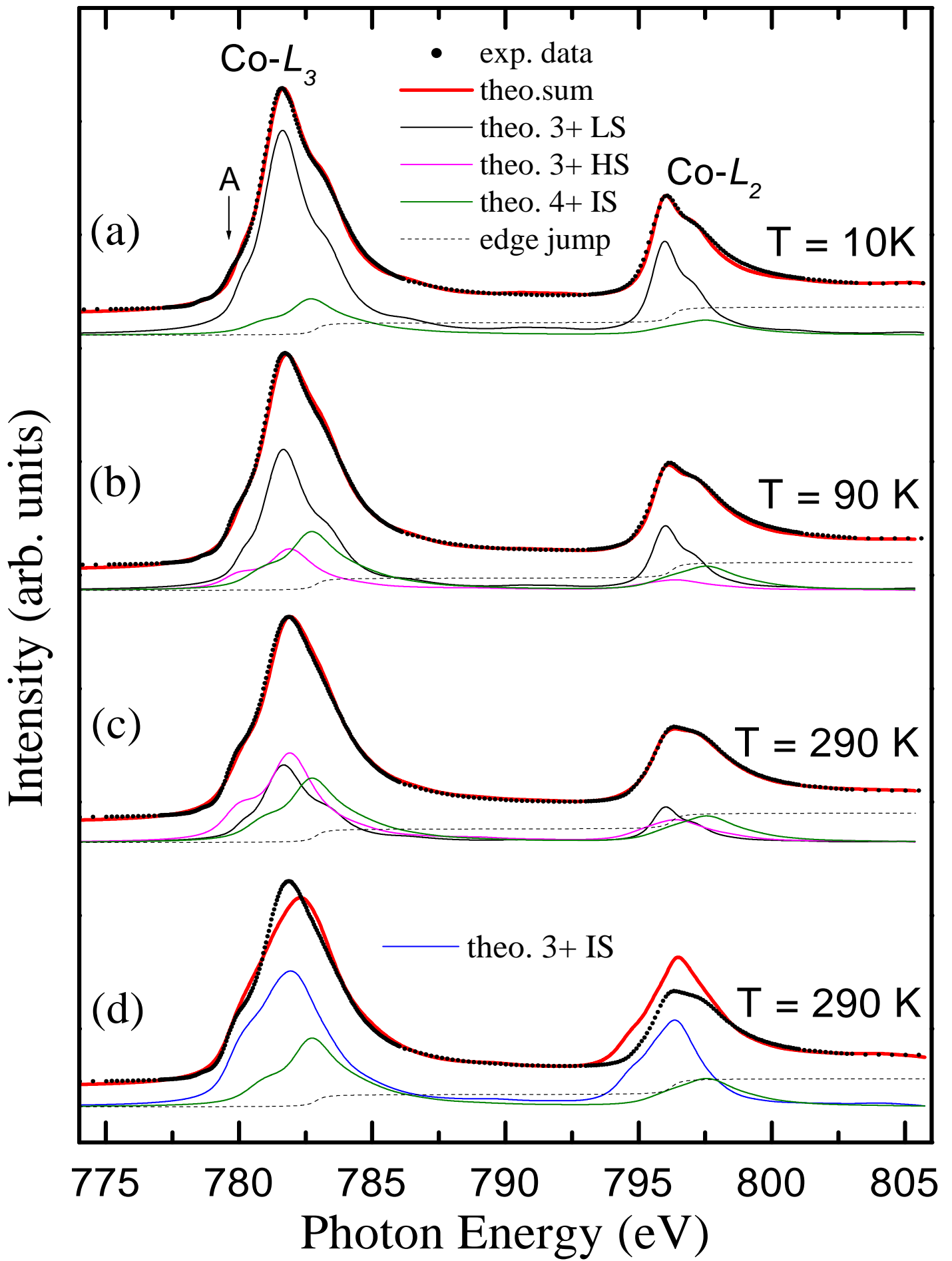

NewFig6
Guillou et al.

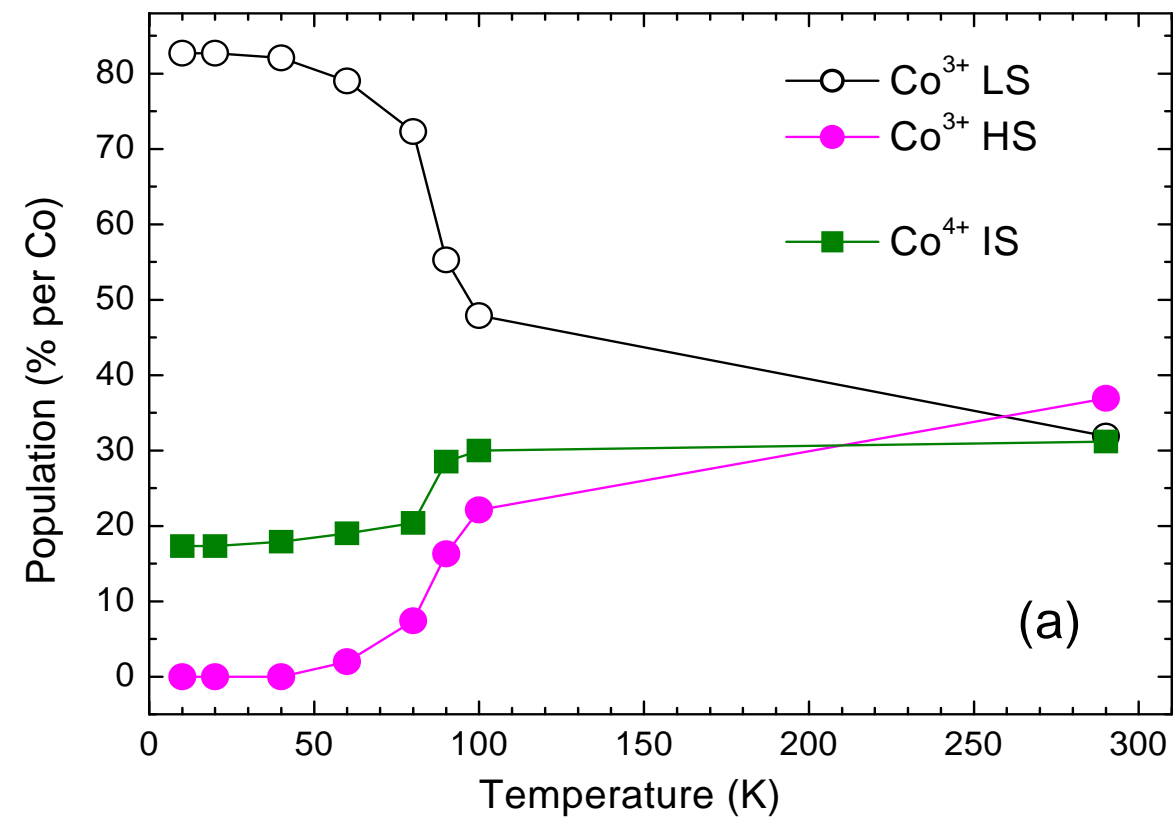

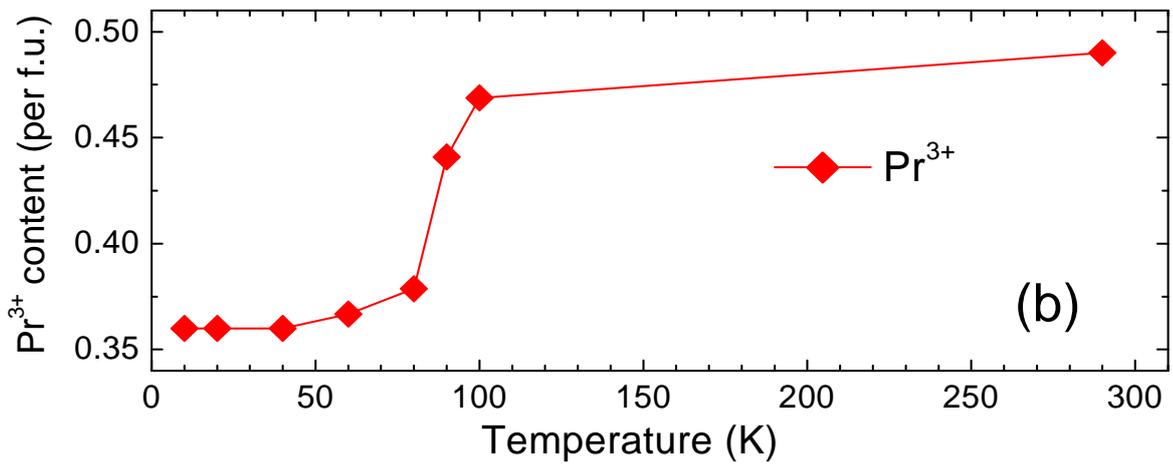

Figure 7